\documentclass[11pt,preprint2]{aastex}

\begin{document}

\title{A Silver Anniversary Observation of the X-Ray Luminous SN 1978K
in NGC 1313}

\author{E. Lenz\altaffilmark{1}, E. M. Schlegel\altaffilmark{2}}

\altaffiltext{1}{Tom C. Clark High School, San Antonio, TX 78249}

\altaffiltext{2}{Department of Physics and Astronomy,
University of Texas-San Antonio, San Antonio, TX 78249}

\begin{abstract}

We describe the results of a 2003 {\it Chandra} ACIS-I observation of
SN 1978K.  The spectrum shows little flux below 0.6 keV, in contrast
to the 2002 {\it Chandra} ACIS-S observation which showed flux to 0.4
keV. Fitting the ACIS-I spectrum alone leads to two solutions
depending upon the value of the column density.  A joint fit using a
dual thermal plasma model applied to the ACIS-I and a contemporaneous
{\it XMM} spectrum, which if fit alone also leads to a two-column
solution, yields a single column density fit.  The fitted temperature
of the joint fit for the soft component remains constant within the
errors from previous {\it Chandra, XMM,} and {\it ASCA} data.  The
hard temperature recovers from its 2000-2002 decline and corresponds
to an increase in the column density during that time.  The hard (2-10
keV) light curve is confirmed to be declining.  The derived number
density represents a lower limit of $\sim$10$^5$ depending upon the
adopted filling factor of the emitting volume, leading to an estimated
mass cooling rate of $\sim$0.1-0.15 M$_{\odot}$ yr$^{-1}$.

\end{abstract}


\section{Introduction}

The late X-ray emission of supernovae remains relatively
unexplored. SN 1987A's initial emission turned on near day 115 and
merged into the background near day 400 post-explosion
\citep{Inoue91}. Recent observations show a reappearance of the
emission as the shock propagates into circumstellar matter
\citep{Zhekov06}. Other than SN 1987A, the only supernovae studied
beyond ${\sim}$500 days have been SN 1978K (\cite{Schlegel04}, and
references therein, hereafter, S04), SN 1995N \citep{Chandra05}, SN
1993J \citep{Zimm03}, and SN 1986J \citep{Temple05} plus a few objects
recovered at late times (eg, SN 1979C and SN 1970G;
\citealt{ImmKuntz05,Immler05}) that presumably will continue to be
studied. Of these, SN 1978K has been the most luminous and hence the
most frequently observed.

Though SN 1978K was first detected as a powerful radio source in 1982
\citep{R93}, it was not realized that it emitted X-rays until an
observation was obtained with the {\it ROSAT} PSPC in 1992. The
subsequent investigation examined archival optical plates, uncovered a
light curve and assigned an explosion date near 1978 May 22
\citep{R93}. Follow-up observations of SN 1978K in the X-ray,
ultraviolet, optical, and radio bands were obtained during the mid-
and late-1990s with {\it ASCA}, the {\it Hubble Space Telescope}, the
{\it ROSAT} High Resolution Imager (HRI), the Australia Telescope
Compact Array (ATCA), and the Anglo-Australian Observatory (AAO)
\citep{S99}.

S04 described a 2002 {\it Chandra} and a 2000 {\it XMM-Newton}
observation of SN 1978K.  The spectra were best fit by a
two-temperature variable-abundance optically thin gas model with
temperatures of 0.6 and $\sim$3 keV.  The authors reported Si emission
in the soft component at 90\% significance.  The flux in the 2-10 keV
band showed the first hint of a decline in the X-ray flux from the
1990s plateau.

Since the 2004 paper, additional data have become available.  Here we
briefly describe a {\it Chandra} ACIS-I observation of SN 1978K
obtained in 2003 October and an {\it XMM-Newton} observation obtained
in 2003 November.  We first describe the data and the spectral fit,
then discuss the results in the context of the previous observations.
Throughout we adopt a distance to NGC 1313 of 4.13$\pm$0.11 Mpc
(M\'endez et al. 2002).

\section{Data and Analysis}

{\it Chandra} observed SN 1978K and the other X-ray emitting sources
in NGC 1313 on 2003 Oct 2 for 14,827 seconds with the ACIS detector
(observation ID 3551, PI G. Garmire) \citep{Garmire03}. The aim point
of the detector fell on the front-illuminated I3 CCD.  The aimpoint on
the I3 CCD differs from previous observations that were obtained with
the back-illuminated S3 CCD; the effective area behaviors of the two
types of CCDs differ.  We discuss our approach to this difference
below.

We extracted a background light curve and spectrum from a region
proximal to the source.  This much larger, source-free region was
1$'$.5 in radius and was positioned immediately west of SN 1978K.  We
examined the light curve for bright transients or flaring events and
did not detect any evidence for either, consequently the good exposure
time was not reduced from 14.8 ksec. The net source count rate was
$\sim$0.083 ${\pm}$ 0.002 counts s$^{-1}$.

We fit the extracted spectra using XSPEC version 12.2.1
\citep{Arnaud96}; for the background spectrum, we adopted a simple
power law continuum with a gaussian at ${\sim}$1.8 keV to model the
possible Si fluorescent feature present in the ACIS background
(S04).

The source events were extracted and binned to a spectrum using a
region at the location of SN 1978K with an aperture of radius
$\sim$18$''$.5.  Though the sharp point-spread function of the {\it
Chandra} mirrors may cause concern about possible event pileup, the
large off-axis angle of 2$'$.6 and low net count rate ensure that
pileup in this case is negligible.  

A single spectral bin exists below 0.6 keV which is consistent with
the background; this is in marked contrast to the 2002 spectrum for
which source events were detected to $\sim$0.4 keV.  We refer the
reader to S04 for a description of a variety of models that did not
provide good fits to that data.  Our adopted best-fit model is
described below and is based on the best-fit model from 2004 paper.

In S04, we used variable-abundance thermal plasma models ('VMEKAL' in
{\tt xspec}, \citealt{Arnaud96}).  In the interim, updated atomic
parameters have become available through the APED project
\citep{Smith03}.  Along with the database, the project has developed a
new thermal plasma model for {\tt xspec} ('VAPEC') that we have used
in fitting the data discussed here.  

\begin{figure} 
\centering 
\caption{The best fit VAPEC model to the {\it Chandra} ACIS I3 2003
spectrum fitting only the ACIS spectrum.  The stepped line indicates
the best-fit VAPEC model.  The lower stepped line and data are the
{\it Chandra} background spectrum and illustrates that the background
is a small fraction of the source below $\sim$5 keV.}
\label{sn_specCh}
%\scalebox{0.35}{\rotatebox{-90}{\includegraphics{sn78k_spec_oct03.ps}}}
\scalebox{0.35}{\rotatebox{-90}{\includegraphics{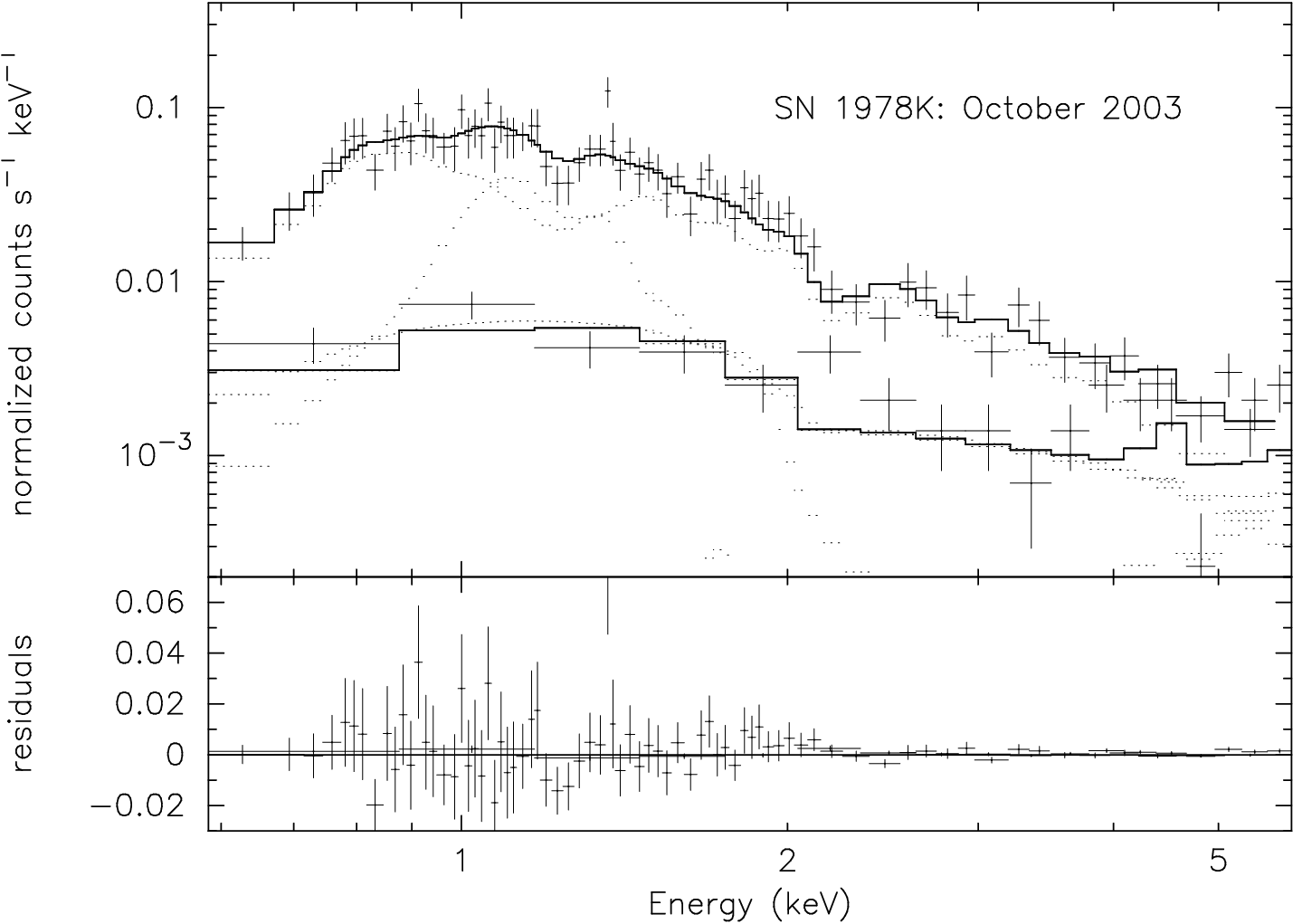}}}
\end{figure}

To fit the 2003 October {\it Chandra} spectrum, we may either fit the
data without regard to previous fits, or constrain the fit assuming
little change over the intervening twelve months.  This consideration
is critical because of the differing spectral responses of the -S and -I
chips.  Given the effective area behavior of the ACIS-I3 CCD, the
critical parameter to constrain is the column density. Fixing its
value at 0.15, the best fit from the 2004 paper, we obtain a fit that
differs largely in flux: the soft and hard temperatures are
essentially identical (Figure~\ref{sn_specCh}, Table~\ref{fit_table},
referred to as VAPEC-L, for Low N$_{\rm H}$).

If we do not place any constraints on the 2003 spectrum, the fit with
the lowest ${\chi}^2/{\nu}$ yields a high column, N$_{\rm H}$
$\sim$0.63$^{+0.18}_{-0.19}{\times}$10$^{22}$ cm$^{-2}$ and a low soft
temperature of $\sim$0.28 keV (Table~\ref{fit_table}, referred to as
VAPEC-H, for High N$_{\rm H}$).  The hard temperature drops slightly
to 2.22 keV but the error bars cover the value from the dual fit.  The
anti-correlation between N$_{\rm H}$ and the soft temperature is
expected given that there are few data points below ${\sim}$0.8 keV to
constrain the parameters separately. We note that the 99\% contour
includes the values obtained from the 2000 and 2002 data sets. We will
include both possibilities in our discussion.

The seemingly abrupt change in the N$_{\rm H}$ parameter could stem
solely from the differing behaviors of the effective areas of the -I
vs -S CCDs which differ, particularly at energies below $\sim$0.7 keV
and above $\sim$4 keV \citep{CPG}.  Consequently, a spectral fit to the
I spectrum could lead to systematic differences from the previous
spectrum.

To investigate any possible systematic differences, we extracted a
spectrum from an {\it XMM-Newton} observation of SN1978K.  The {\it
XMM-Newton} observation was obtained on 2003 November 25 (obsid
0150280101, PI I. Smith) for $\sim$12 ksec.  The spectrum was
extracted after filtering the data to eliminate times of high
background that generally afflict {\it XMM} observations.  The
extracted net spectrum contains $\sim$600 counts.  The background
spectrum was obtained from 1$'$.6 south of SN1978K's location.
It was featureless in the area of interest ($\sim$0.5-3 keV) and
represents $<$15\% of the source signal.

\begin{figure} 
\centering 
%\scalebox{0.35}{\rotatebox{-90}{\includegraphics{xmm_T1T2_cont.ps}}}
\scalebox{0.35}{\rotatebox{-90}{\includegraphics{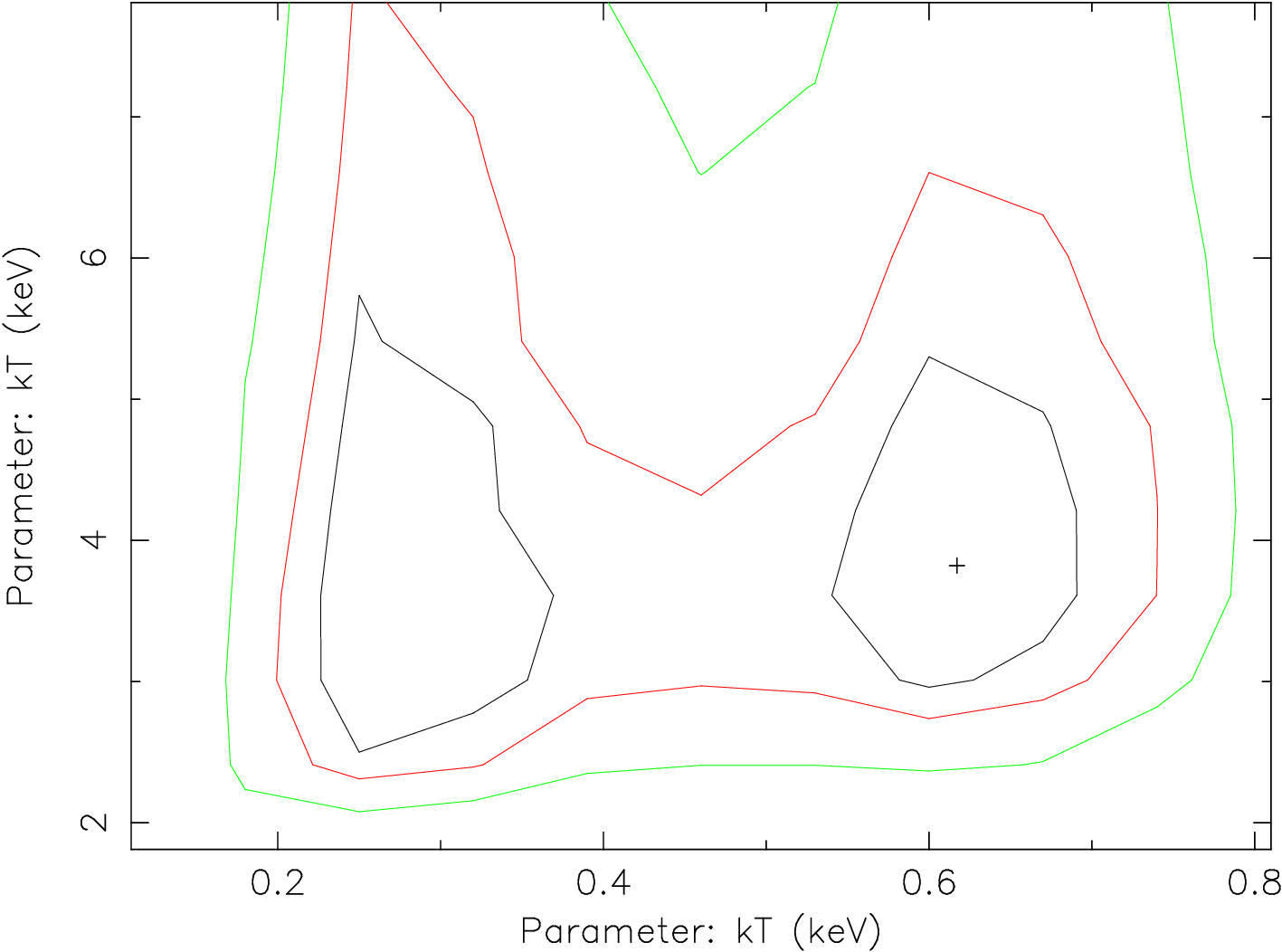}}}
\caption{Contour plots of the fitted temperatures for the {\it XMM}
November 2003 spectrum illustrating the possible double-valued soft temperature.}
\label{xmm_Tcont}
\end{figure}

In fitting the VAPEC model to the {\it XMM} spectrum, we again obtain
a dual solution to the soft temperature because of differing values of
the column density (Figure~\ref{xmm_Tcont}).  Otherwise, the {\it
Chandra}-only and {\it XMM}-only fits are similar.

\begin{figure} 
\centering 
\caption{The best fit VAPEC model to the {\it Chandra} ACIS I3 and
{\it XMM} MOS-1 2003 spectra.  The background spectra have been
suppressed for clarity.  The upper curve is {\it Chandra}; the lower
is {\it XMM}.  (See the online journal for a color version of this
figure.)}
\label{sn_spec_both}
%\scalebox{0.35}{\rotatebox{-90}{\includegraphics{sn78k_X1C1_spec_v2.ps}}}
\scalebox{0.35}{\rotatebox{-90}{\includegraphics{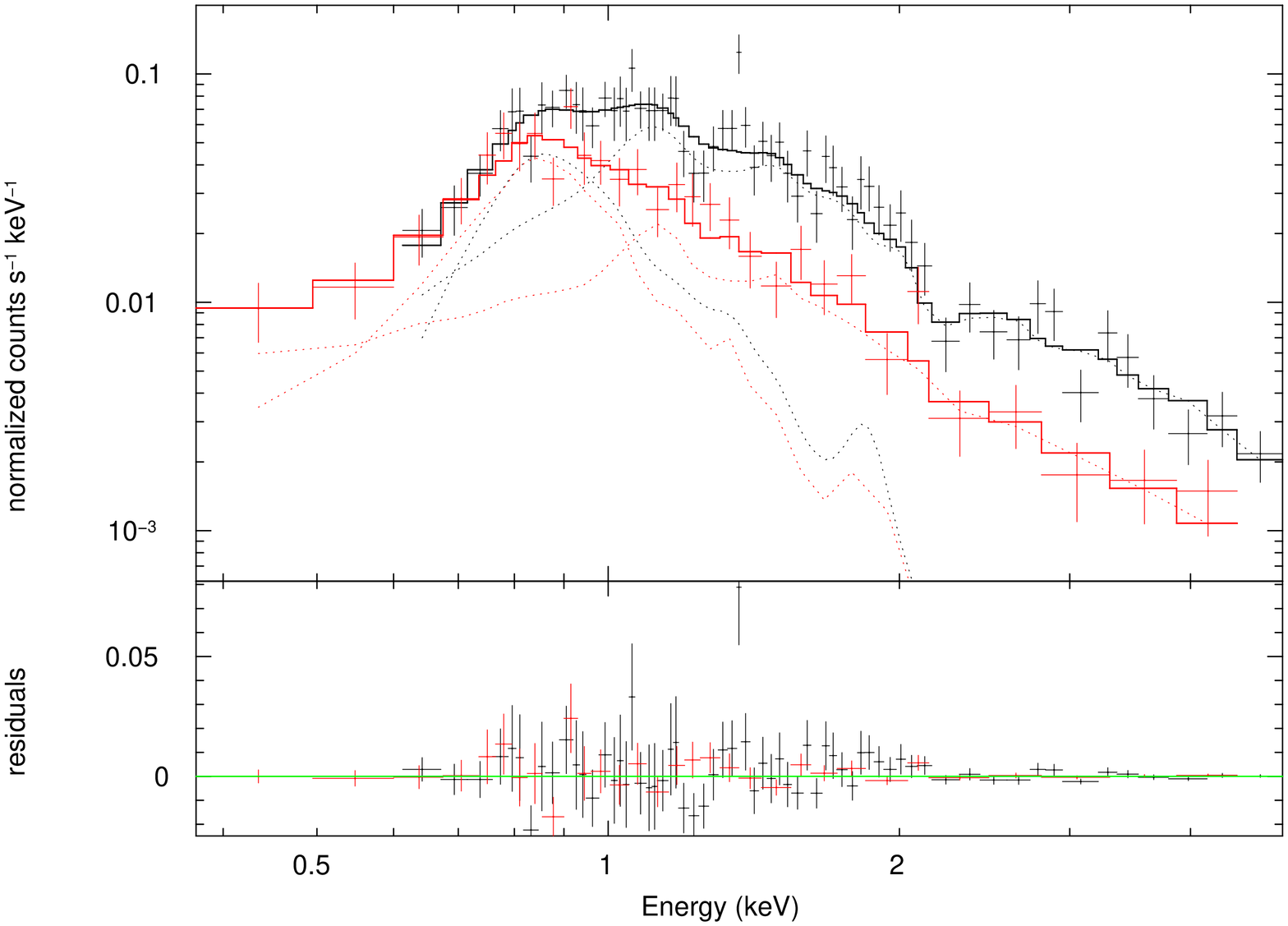}}}
\end{figure}

\begin{figure} 
\centering 
%\scalebox{0.35}{\rotatebox{-90}{\includegraphics{sn78k_X1C1_NHT1_cont.ps}}}
%\scalebox{0.35}{\rotatebox{-90}{\includegraphics{sn78k_X1C1_NHT2_cont.ps}}}
\scalebox{0.35}{\rotatebox{-90}{\includegraphics{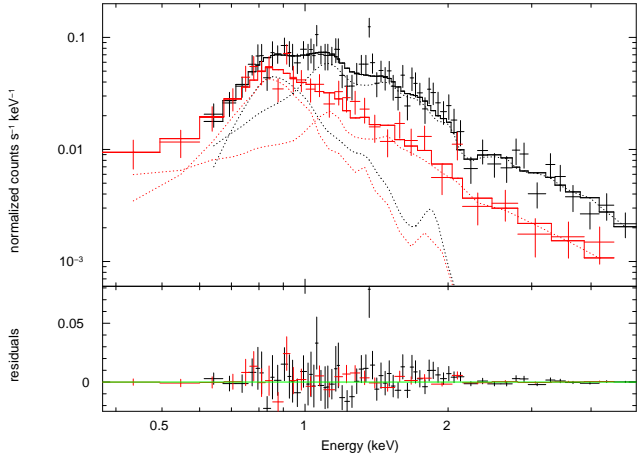}}}
\scalebox{0.35}{\rotatebox{-90}{\includegraphics{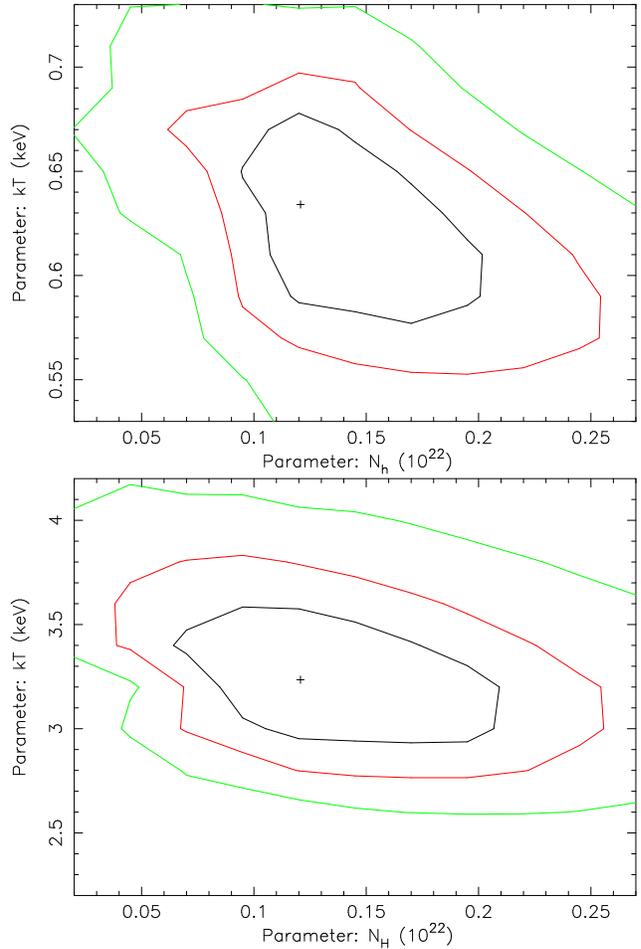}}}
\caption{Contour plots of the fitted temperatures and column density
for the {\it XMM} November and {\it Chandra} 2003 spectra. (top) soft
component temperature and column density; (bottom) hard component
temperature and column density.}
\label{T1T2_cont}
\end{figure}

To extract the best spectral fit parameters with the least number of
constraints, we fit the ACIS-I3 and {\it XMM} data simultaneously.
Figure~\ref{sn_spec_both} shows the resulting spectral fit;
Figure~\ref{T1T2_cont} shows the temperature contours.  The
simultaneous fit yields a lower value for the column density than
described in S04 (0.12$^{+0.07}_{-0.05}$ vs. 0.23$^{+0.04}_{-0.03}$ in
units of 10$^{22}$ cm$^{-2}$), an identical soft temperature
(0.64$^{+0.08}_{-0.05}$ keV vs. 0.61$^{+0.04}_{-0.05}$ keV) and a
higher hard temperature (3.36$^{+0.72}_{-0.38}$ keV
vs. 3.16$^{+0.44}_{-0.40}$ keV), although the 90\% confidence ranges
of the hard temperatures overlap.

 Variable models such as VMEKAL and VAPEC permit altering the
abundances of elements known to produce emission lines. We
investigated elemental abundances because S04 reported Si to be
present in the soft spectral component.  We started with all element
abundances frozen at 1.0 and checked the spectrum for visible lines.
We in turn relaxed and fixed each element's abundance, each time
looking for significant changes in the ${\chi}^2$/${\nu}$.  In all
cases, including Si, within the errors, the abundances were consistent
with 1.0 (Figure~\ref{sn_spec_both}).  We report upper limits only for
Si (Table~\ref{fit_table}).

\section{Discussion} 

We base our discussion on the dual {\it Chandra-XMM} spectral fit
because the evidence shows that fitting only one spectrum leads to
systematic differences with the prior {\it Chandra} ACIS-S spectra
(S04).

The normalization values in the hard and soft components for the 2003
data (Table~\ref{fit_table}) are nonzero at the 90\% level, so it is
evident that the two components remain significantly detected as was
true in the 2002 {\it Chandra} observation.

To maintain consistency between the 2003 {\it Chandra} data and the
earlier {\it XMM-Newton}, {\it ASCA}, and 2002 {\it Chandra} data, we
also refit the previously-published spectra using the dual-component
VAPEC model. The fits yielded values very similar to those of the
published VMEKAL values. The updated numbers are shown in
Table~\ref{APEC_table} and are included for completeness.  Differences
between column densities and temperatures were generally within 5\% of
the values published in S04.  The {\it ASCA}-1 and {\it ASCA}-2 values
appeared to differ the most ($\sim$10-20\%), likely due to the poor
low-energy response.\footnote{For additional information, see the {\it
ASCA} Guest Observer Facility's "Watch Out" page
(heasarc.gsfc.nasa.gov/docs/asca/watchout.html).}

We include in Table~\ref{fluxes} the fluxes for the {\it Chandra} and
{\it XMM}-only model fits, the joint {\it Chandra-XMM} fit, as well as
the re-fitted spectra from S04.  Of particular interest is the soft
{\it Chandra} component that shows a high column density.  The
corresponding unabsorbed flux is very high, demonstrating the
necessity for care in handling the ACIS-I spectrum.

\begin{figure} 
\centering 
%\scalebox{0.35}{\rotatebox{-90}{\includegraphics{sn78k_lc_052.ps}}}
\scalebox{0.35}{\rotatebox{-90}{\includegraphics{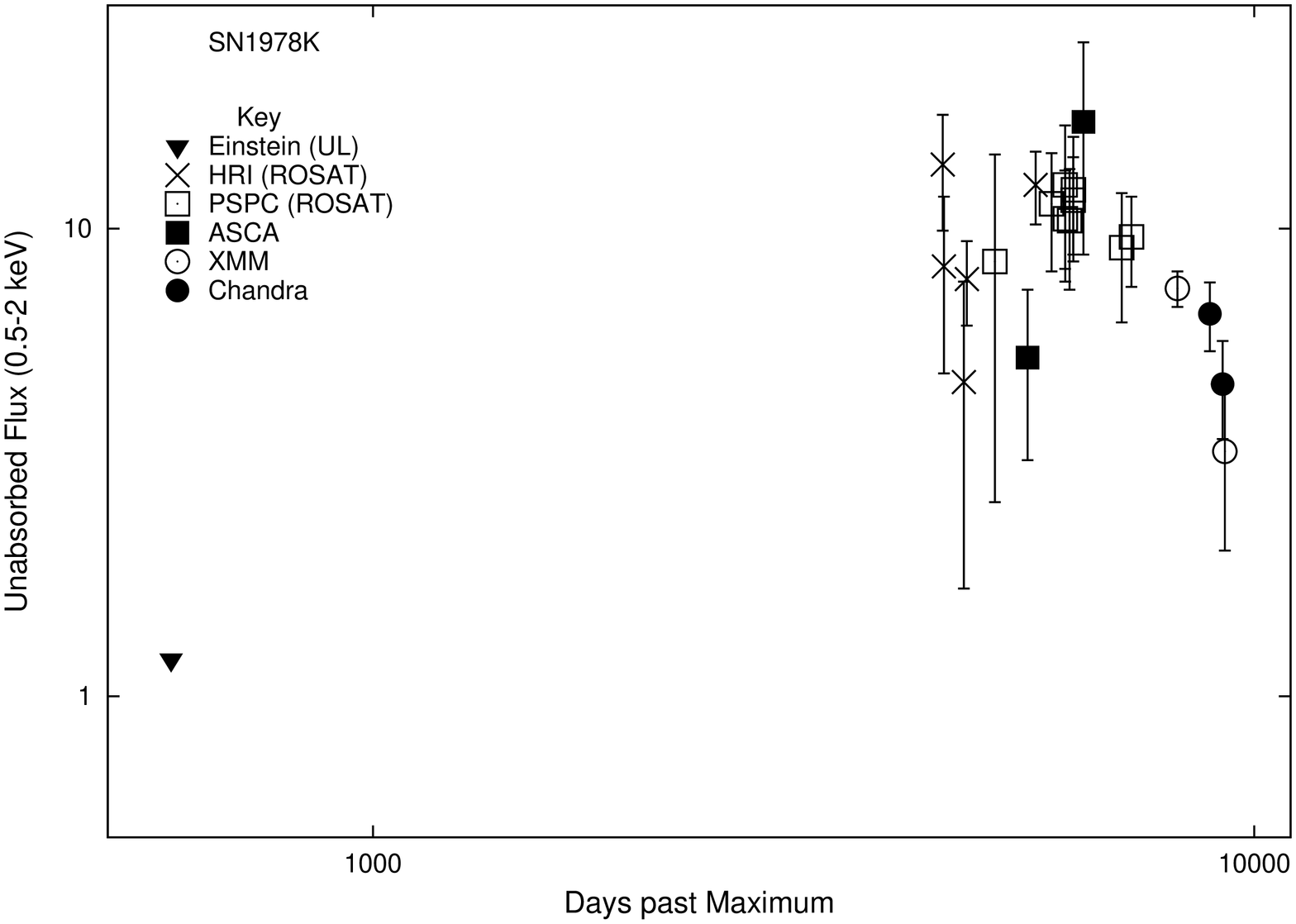}}}
%\scalebox{0.35}{\rotatebox{-90}{\includegraphics{sn78k_lc_2_10.ps}}}
\scalebox{0.35}{\rotatebox{-90}{\includegraphics{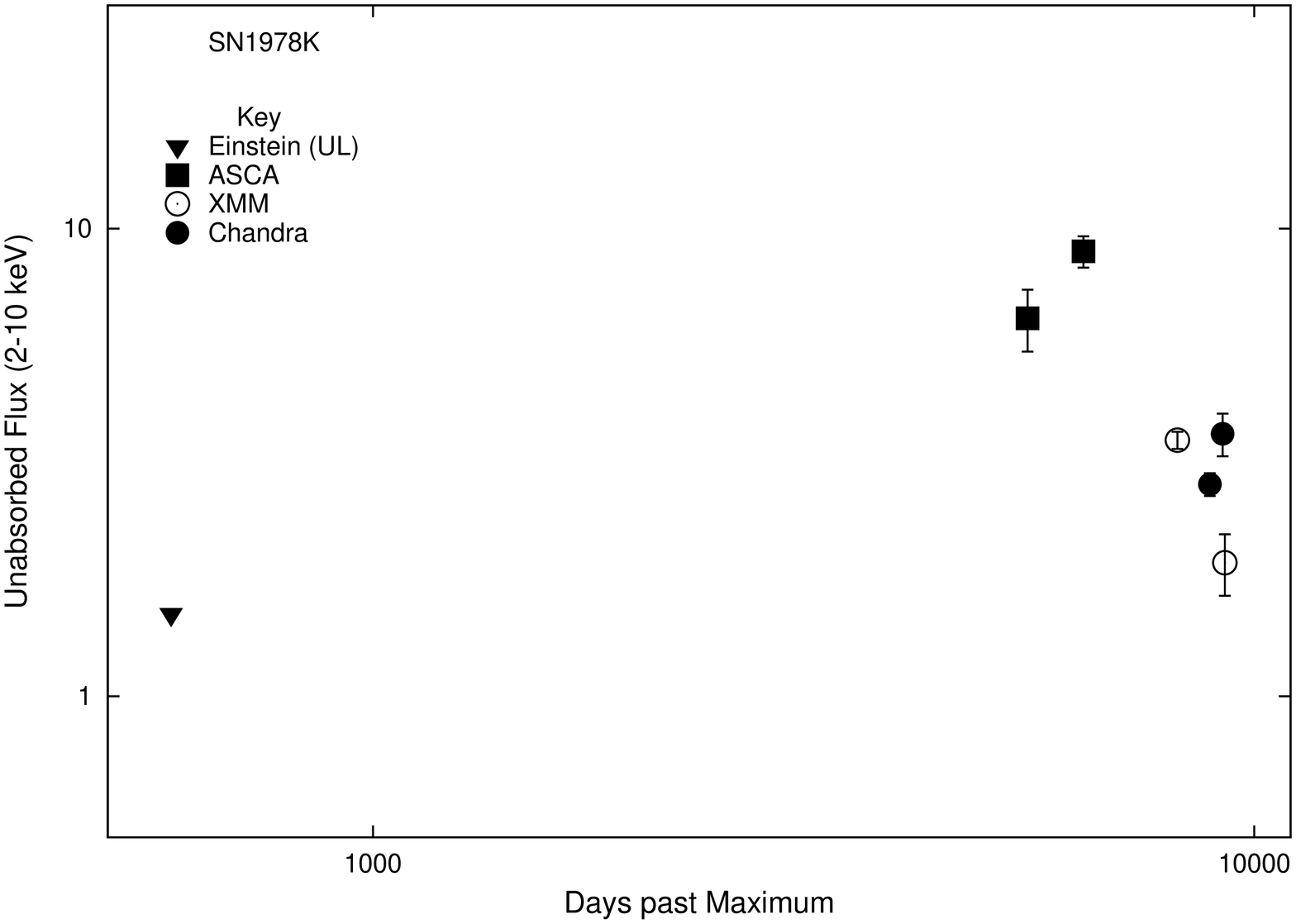}}}
\caption{Light curves of (top) soft and (bottom) hard spectral
components.  The fluxes used in all cases are the {\it un}absorbed
values.}
\label{lc}
\end{figure}

With a consistent set of model fits, the light curves of the soft and
hard components may be constructed.  Figure~\ref{lc} displays the soft
(top) and hard (bottom) {\it un}absorbed light curves.  Both light
curves shows a continuing decrease in flux, confirming the decline
reported in S04.

\begin{figure}
\centering
\caption{Schematic plot of the fitted parameter values vs. time
including N$_{H}$, kT$_{\rm soft}$, kT$_{\rm hard}$ and Si abundance.
The kT$_{\rm hard}$ and Si abundance values have been divided by a
constant to fit them into the range.  Several data points have been 
shifted in age for clarity.}
\label{param_plot}
%\scalebox{0.35}{\rotatebox{-90}{\includegraphics{sn78k_params.ps}}}
\scalebox{0.35}{\rotatebox{-90}{\includegraphics{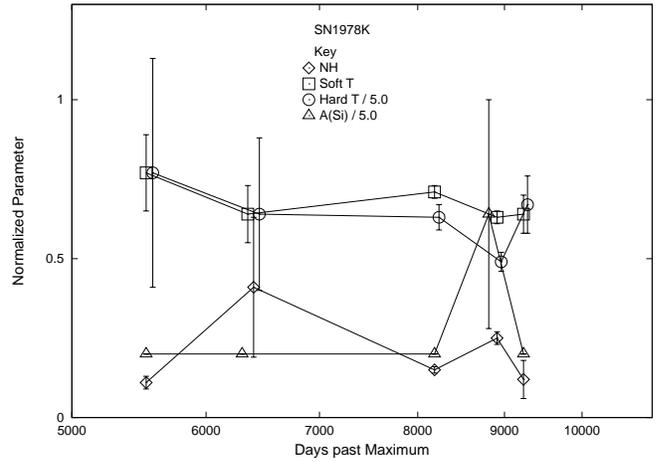}}}
\end{figure}

Though the 2002 and 2003 data share the dual-component structure,
significant differences exist between the two epochs.  The results are
shown schematically in Figure~\ref{param_plot}.  Note that the values
of two of the parameters have been divided by a constant to place the
points into the plot.

The soft component kT has held steady within the errors at $\sim$0.6
keV.  In contrast, the hard component temperature has held steady at
$\sim$3.2-3.4 keV with the exception of the day 8910 spectrum, when it
dropped to $\sim$2.4 keV at the same time that the column density
increased.

The column density varied from $\sim$0.15 to 0.25 (in units of
10$^{22}$ cm$^{-2}$) and back.  The known column toward NGC 1313 is
$\sim$3.7${\times}$10$^{20}$ cm$^{-2}$ (S04).  As pointed out in S04,
the E$_{\rm B - V}$ value from the optical spectrum is $\sim$0.31
which corresponds to N$_{\rm H}$ $\sim$1.6${\times}$10$^{21}$
cm$^{-2}$, a value matched by the fitted N$_{\rm H}$ for days 5531,
8184, and 9209/9263.  The E$_{\rm B-V}$-derived value also matches the
day 6401 column within the errors.  The difference between the known
column to NGC 1313 and the fitted or derived column values requires
considerable absorption local to SN1978K.  That the column density
increased at day 8910 and then declined suggests the existence of a
recent, local density enhancement.

Finally, another difference in the two data sets is the abundance of
Si. The 2003 spectra show no detection of Si in either component,
though the 2002 data showed an abundance well above solar and
significant at the 90\% level.

The ratio of kT$_{\rm soft}$ and kT$_{\rm hard}$ determines the power
law indices, ({\it n} and {\it s}), of the ejecta and circumstellar
matter distribution, respectively \citep{Fransson96}. The usual
adoption of {\it s} = 2 allows us to solve for {\it n} using one of
the predictions of the thin shell model \citep{Fransson96}:
(T$_r$/T$_f$) = ((3--s)$^2$/(n--3)$^2$) where T$_r$ is the reverse
shock temperature and T$_f$ is the forward shock temperature.
{\it Assuming} that T$_r$ is given by T$_{soft}$ and T$_f$ is given by
T$_{hard}$, {\it n} is calculated as 5.29${\pm}$0.88 for the 2003 {\it
Chandra} observation. The {\it n} values for the 2002 {\it Chandra},
{\it XMM-Newton}, {\it ASCA}-2 and {\it ASCA}-1 observations are
4.96${\pm}$0.04, 5.29${\pm}$0.10, 7.97${\pm}$0.62 and 5.23${\pm}$0.83,
respectively. The weighted mean value is 5.02.  This is a lower value
for the power law index than is generally assumed for mass
distributions (typically 8-12; \citealt{Fransson96}), but within the
errors of other X-ray-emitting SNe (eg, SN1993J, S04).  Smaller values
of the index lead to longer cooling times as is observed in SN1978K.

We may infer X-ray properties following the arguments in \cite{SP06}
that were based on analyses first used by Immler and co-workers
(summarized in \cite{IL03}).  Thermal plasma emission is given by
L$_{\rm X}$ = ${\Lambda}(T)n^2_{\rm e}V_{\rm X}$ where ${\Lambda}$(T)
is the emissivity, n$_{\rm e}$ is the electron density, and V$_{\rm
X}$ is the emitting volume.  ${\Lambda}$ is $\sim$10$^{-23}$ erg
s$^{-1}$ cm$^{-3}$ for plasmas with temperatures of $\sim$0.5-10 keV
in the 0.2-5 keV band \citep{Raymond76}.  The quantity n$_{\rm
e}V_{\rm X}$ is the volume emission measure.  An estimate of the
volume then delivers an estimate of the number density from which we
may calculate the mass of the cooling gas, M$_{\rm X}$
(Table~\ref{inferred}).

Table~\ref{inferred} lists the inferred quantities where we also
include a volume filling factor, ${\phi}$, that describes the emitting
matter within the volume enclosed by the shock.  The shock volume is
estimated from the maximum observed line width of the 1996 optical
spectrum ($\sim$600 km s$^{-1}$, \citealt{S99}).  To estimate
velocities for the other epochs, we adopt the velocity profile from SN
1988Z, t$^{-(5/7)}$, as described by \citep{Aretxaga99}. We justify
the adoption of this velocity profile on the basis of the similar
X-ray luminosities and optical spectral line widths.  SN1988Z is the
only other supernova with sufficient data at late times, hence our
adoption of the velocity profile.  

If the plasma uniformly fills the shock volume, the density must be
$>$10$^5$ cm$^{-3}$.  An order of magnitude higher density requires a
filling factor smaller by the factor ${\sqrt{10}}$.  These estimates
require the assumption of collisional ionization equilibrium which
occurs if the product $n_{\rm e}t > 10^{13}$ s cm$^{-3}$.  Our
estimates at the age of SN 1978K exceed that criterion by at least a
factor of 20.

If we take at face value the results of the spectral fits, then the
column density increased between 2000 and 2003 from
${\sim}$1.5$\times$10$^{21}$ cm$^{-2}$ to ${\sim}$2.5$\times$10$^{21}$
cm$^{-2}$ and back. Is the increase real?  The larger effective area
of the ACIS-S CCD in the 2002 observation, as well as the longer
exposure time, serve to establish that the increase in column density
is not an instrumental effect or an artifact of effective area.

Is the increase physically reasonable?  The path length during that
interval increased by ${\sim}$3$\times$10$^{15}$ cm based on the
adopted shock velocity from Table~\ref{inferred}.  This translates to
an increase in the number density from ${\sim}$2--4$\times$10$^5$
cm$^{-3}$ to ${\sim}$7-8$\times$10$^5$ cm$^{-3}$, or an increase of a
factor of ${\sim}$1.7--2. This range has been observed in SN 2001ig
where an increase in the radio flux by a factor of ${\sim}$3 was
attributed to an increase in the number density by a factor of
${\sim}$2 over a time span of ${\sim}$150 d \citep{Ryder04}.  For
SN1978K, the time span could be as long as 400-500 days, assuming the
enhancement started immediately after the {\it XMM} observation in
2000 and halted immediately before the {\it Chandra} observation of
2003.  Hence we conclude that the increase in column density is
physically possible.

That the Si abundance also appears to have increased implies enhanced
emission as a shock overran a density enhancement and specifically a
Si enhancement.  A Si enhancement requires production of Si which
occurs in stars of mass 11-35 M$_{\odot}$ based on a study of the
integrated yield assuming a Salpeter initial mass function
\citep{LC06}.

In summary, with the 2003 observation, we now see a higher column
occurring approximately in 2002.  The increase in column density can
be explained as an local increase in prior mass loss as observed in
several X-ray-emitting SNe (eg, SN1979C, SN2001ig).  This behavior
argues for increased time sampling of the late phases of
X-ray-emitting SNe.  The benefit of such observations is the
possibility of insight into the mass loss of massive stars.  Will, for
example, the flux from SN1978K increase if the outgoing shock runs
over another density enhancement?  We also confirm the decrease in the
X-ray flux from SN1978K that started in $\sim$2000-2002 and was first
reported in S04, a forlorn epitaph for a silver anniversary.

\section{Acknowledgements}

We thank the referee for comments that improved this paper. 
The research of EL and EMS was supported in part by Grant GO4-5017A from the
{\it Chandra} X-ray Center operated by the Smithsonian Astrophysical
Observatory under contract to NASA.

\newpage

\begin{table*}
\centering
\caption{Results of Spectral Fits\tablenotemark{a}}
\label{fit_table}
\begin{tabular}{lrrrrrrrr}
       &                   &        &  N$_{H}$              &  T$_{\rm soft}$  & T$_{\rm hard}$  &        &        \\
Model  &  ${\chi}_{\nu}^2$ & DoF    & (10$^{22}$ cm$^{-2}$  &  (keV) & (keV)  & Si$_1$, Si$_2$ & Norm-1 & Norm-2 \\ \hline
\multicolumn{9}{c}{October 2003 Chandra Spectrum only} \\
VMEKAL & 1.220 & 87 & 0.72$^{+0.15}_{-0.26}$ & 0.22$^{+0.09}_{-0.05}$ & 1.97$^{+0.53}_{-0.33}$ & 5.3$^{+3.9}_{-2.2}$, $<$54  & 3.62$^{+17}_{-2.20}$E-3 & 49$^{+7.6}_{-7.5}$E-5 \\
%Brems & 1.33 & 83 & 0.93$^{+0.30}_{-0.29}$ & 0.12$^{+0.04}_{-0.02}$ & 1.23$^{+0.41}_{-0.21}$ & 2.72$^{+98}_{-2}$ & 5.97$^{+4.38}_{-2.58}$E-4 \\
VAPEC-L & 1.22 & 72 & 0.15f & 0.63$^{+0.11}_{-0.09}$ & 2.61$^{+0.72}_{-0.38}$ & 1.0, 3.19$^{+2.27}_{-1.60}$ & 7.90$^{+1.12}_{-2.10}$E-5 & 3.32$^{+0.45}_{-0.41}$E-4 \\
VAPEC-H & 1.17 & 71 & 0.63$^{+0.18}_{-0.19}$ & 0.28$^{+0.10}_{-0.06}$ & 2.22$^{+1.12}_{-0.37}$ & 1.0f, 2.75$^{+2.91}_{-0.88}$ & 1.58$^{+0.28}_{-0.23}$E-3 & 4.03$^{+0.34}_{-0.32}$E-4 \\
 \\
\multicolumn{9}{c}{November 2003 XMM Spectrum only} \\
VAPEC  & 0.89 & 29 & 0.15$^{+0.08}_{-0.05}$ & 0.62$^{+0.12}_{-0.42}$ & 3.83$^{+4.2}_{-1.6}$ & $<$3.5, $<$3.0 & 8.91$^{+}_{-}$E-5 & 22.89$^{+}_{-}$E-5 \\
 \\
\multicolumn{9}{c}{Dual Chandra/XMM Spectra} \\
VAPEC & 0.98 & 95 & 0.12$^{+0.07}_{-0.05}$ & 0.64$^{+0.08}_{-0.05}$ & 3.36$^{+0.53}_{-0.39}$ & $<$4.5, $<$3.3 & 6.71$^{+2.84}_{-2.28}$E-5 & 43.62$^{+3.39}_{-3.17}$E-5 \\
      &      &    &                        &                        &                        &                & 7.18$^{+3.44}_{-2.13}$E-5 & 23.38$^{+3.64}_{-3.18}$E-5 \\
\hline
\end{tabular}
\tablenotetext{a}{Errors are 90\%; a small 'f' indicates a quantity fixed at the specified value.}
\end{table*}

\begin{table*}
\centering
\caption{APEC Fits to Previous CCD-Resolution Spectra\tablenotemark{a}}
\label{APEC_table}
\begin{tabular}{lrrrr}
Observation      & Age\tablenotemark{b}   & N$_{H}$         & T$_{soft}$      & T$_{hard}$ \\  \hline
{\it Chandra}    & 8910      & 0.25${\pm}$0.02 & 0.63${\pm}$0.02 & 2.43${\pm}$0.16 \\
{\it XMM-Newton} & 8184      & 0.15${\pm}$0.01 & 0.71${\pm}$0.02 & 3.17${\pm}$0.21 \\
{\it ASCA}-2     & 6401      & 0.41${\pm}$0.22 & 0.64${\pm}$0.09 & 3.19${\pm}$1.18 \\
{\it ASCA}-1     & 5531      & 0.11${\pm}$0.02 & 0.77${\pm}$0.12 & 3.83${\pm}$1.80 \\ \hline
\end{tabular}
\tablenotetext{a}{Details of the previous observations are listed in
\cite{Schlegel04}. Units for the columns are: Age: days; N$_{\rm H}$:
10$^{22}$ cm$^{-2}$; T$_{soft}$ and T$_{hard}$: keV.}
\tablenotetext{b}{Age based on date of optical maximum = 1978 May 22, MJD 43650.}
\end{table*}

\begin{table*}
\centering
\caption{Fluxes\tablenotemark{a} for VAPEC model\tablenotemark{b}: {\it Chandra}, {\it XMM-Newton}, and {\it ASCA}}
\label{fluxes}
\begin{tabular}{lrrrrrrrr}
        & Age & Band & \multicolumn{2}{c}{Complete}  & \multicolumn{2}{c}{Soft} & \multicolumn{2}{c}{Hard} \\
Observation & (days) & (keV) & Abs'd & Unabs'd & Abs'd & Unabs'd & Abs'd & Unabs'd  \\  \hline
joint {\it Chandra} & 9209 & 0.5-2 & 3.33 & 4.65 & 1.06 & 1.59 & 2.27 & 3.07 \\
                    &      &  2-10 & 3.58 & 3.64 & 0.04 & 0.05 & 3.54 & 3.59 \\
 + {\it XMM}        & 9263 & 0.5-2 & 2.35 & 3.34 & 1.13 & 1.70 & 1.61 & 1.92 \\
                    &      &  2-10 & 1.90 & 1.93 & 0.05 & 0.05 & 1.85 & 1.88 \\  \hline
{\it XMM} only     & 9263  & 0.5-2 & 2.36 & 3.72 & 1.22 & 2.09 & 1.13 & 1.63 \\
                   &       & 2-10  & 2.20 & 2.24 & 0.05 & 0.05 & 2.15 & 2.18 \\
{\it Chandra} only & 9209  & 0.5-2 & 3.29 & 4.86 & 1.56 & 2.28 & 2.24 & 3.00 \\
 Low N$_{\rm H}$   &       &  2-10 & 3.58 & 3.64 & 1.10 & 1.11 & 3.51 & 3.56 \\
{\it Chandra} only & 9209  & 0.5-2 & 3.30 & 27.2 & 2.25 & 23.8 & 1.46 & 3.76 \\
 High N$_{\rm H}$  &       &  2-10 & 3.41 & 3.63 & 1.15 & 1.12 & 3.35 & 3.60 \\  \hline
2002 {\it Chandra}\tablenotemark{c} & 8910  & 0.5-2 & 4.25 & 6.57 & 2.06 & 3.37 & 2.21 & 3.20 \\
                   &       &  2-10 & 2.78 & 2.84 & 0.11 & 0.12 & 2.67 & 2.72 \\
{\it XMM-Newton}\tablenotemark{c}   & 8184  & 0.5-2 & 4.50 & 7.45 & 2.15 & 3.79 & 2.35 & 3.65 \\
                   &       &  2-10 & 3.44 & 3.53 & 0.17 & 0.18 & 3.27 & 3.35 \\
{\it ASCA}-2\tablenotemark{c}       & 6401  & 0.5-2 & 6.29 & 16.9 & 4.08 & 11.8 & 3.40 & 6.23 \\
                   &       &  2-10 & 8.62 & 8.94 & 1.03 & 1.06 & 8.35 & 8.67 \\
{\it ASCA}-1\tablenotemark{c}       & 5531  & 0.5-2 & 4.25 & 5.30 & 1.02 & 1.31 & 3.21 & 3.96 \\
                   &       &  2-10 & 6.37 & 6.43 & 0.08 & 0.09 & 6.29 & 6.34 \\  \hline
\end{tabular}
\tablenotetext{a}{All fluxes in units of 10$^{-13}$ ergs s$^{-1}$
  cm$^{-2}$.  The values generally differ by a few percent from those
  reported in \cite{Schlegel04}.  We include the values here to be
  complete, having re-defined the best-fit adopted model from that
  paper.}
\tablenotetext{b}{"Abs'd" and "Unabs'd" represent fluxes determined
from the model with (Absorbed) and without (Unabsorbed) the column
density component.}
\tablenotetext{c}{Fluxes recalculated for VAPEC model; values should be 
considered to replace those presented in S04.}
\end{table*}

\newpage

%\newpage

%\newpage

\newpage

%\newpage

\newpage

\newpage

\begin{table*}[h!]
\centering
\caption{Inferred Properties of SN1978K}
\label{inferred}
\begin{tabular}{rrrrrrr}
Age  &  Velocity\tablenotemark{a} & Radius & n$^2_e {\phi} V$\tablenotemark{b} & n$_e {\phi}^{1/2}$ &             & M$_{\rm X}$\tablenotemark{d} \\
(days) & (km s$^{-1}$) & (pc) & (10$^{61}$ cm$^3$) & (10$^5$ cm$^{-3}$) & ${\phi}_7$\tablenotemark{c} & (M$_{\odot}$) \\  \hline
\multicolumn{7}{c}{Soft Component} \\
5531 & 690 & 0.010 & 2.84 & 4.82 & 4.8$\times$10$^{-2}$ & 0.05 \\
6401 & 620 & 0.011 & 26.11 & 12.66 & 1.3$\times$10$^{-1}$ & 0.17 \\
8184 & 520 & 0.011 & 8.06 & 7.03 & 7.0$\times$10$^{-2}$ & 0.10 \\
8910 & 490 & 0.012 & 7.08 & 5.79 & 5.8$\times$10$^{-2}$ & 0.10 \\
9209 & 480 & 0.012 & 3.33 & 3.97 & 3.9$\times$10$^{-2}$ & 0.07 \\
9263 & 475 & 0.012 & 3.55 & 4.10 & 4.1$\times$10$^{-2}$ & 0.07 \\
\multicolumn{7}{c}{Hard Component} \\
5531 & 690 & 0.010 & 20.91 & 13.09 & 1.3$\times$10$^{-1}$ & 0.13 \\
6401 & 620 & 0.011 & 30.25 & 13.62 & 1.4$\times$10$^{-1}$ & 0.18 \\
8184 & 520 & 0.011 & 14.21 & 9.34 & 9.3$\times$10$^{-2}$ & 0.13 \\
8910 & 490 & 0.012 & 12.02 & 7.55 & 7.5$\times$10$^{-2}$ & 0.13 \\
9209 & 480 & 0.012 & 13.52 & 8.00 & 8.0$\times$10$^{-2}$ & 0.14 \\
9263 & 475 & 0.012 &  7.71 & 6.04 & 6.0$\times$10$^{-2}$ & 0.11 \\ \hline
\end{tabular}
\tablenotetext{a}{Velocity estimated adopting velocity profile of t$^{-(5/7)}$ from \cite{Aretxaga99}
 and based upon the optical emission line width (described in text).}
\tablenotetext{b}{${\phi}$ is the dimensionless filling factor; n$^2_eV$ is the emission measure; an
estimate of the volume V leads to an estimate of the number density n$_e$.}
\tablenotetext{c}{${\phi}_7$ assumes n$_{\rm e}$ = 10$^7$ cm$^{-3}$.}
\tablenotetext{d}{M$_{\rm X}$ is the mass of cooling gas in M$_{\odot}$.}
\end{table*}

\newpage

\end{document}